\documentclass[referee]{mn2e}
\usepackage{times,epsfig}

\newcommand{\aap}{A\&A}

\newcommand{\aj}{AJ}
\newcommand{\apj}{ApJ}
\newcommand{\apjl}{ApJL}

\newcommand{\mnras}{MNRAS}

\def \kms{\ifmmode{~{\rm km\,s}^{-1}}\else{~km~s$^{-1}$}\fi}
\def \vhel{\ifmmode{V_{{\rm hel}}}\else{$V_{{\rm hel}}$}\fi}
\def \vsys{\ifmmode{V_{{\rm sys}}}\else{$V_{{\rm sys}}$}\fi}
\def \vobs{\ifmmode{V_{{\rm obs}}}\else{$V_{{\rm obs}}$}\fi}
\def \degree{\ifmmode{^{\circ}}\else{$^{\circ}$}\fi}
\def \lsun{\ifmmode{{\rm\ L}_\odot}\else{${\rm\ L}_\odot $}\fi}
\def \msun{\ifmmode{{\rm\ M}_\odot}\else{${\rm\ M}_\odot$}\fi}
\def \myr{\ifmmode{{\rm\ M}_\odot{\rm\ yr}^{-1}}\else{${\rm\ M}_\odot$ 
yr$^{-1}$}\fi}
\def \teff{\ifmmode{{\rm{T}}_{\rm eff}}\else{${\rm{T}}_{\rm eff}$}\fi}
\def \mdot{\ifmmode{{\rm\dot{M}}}\else{${\rm\dot{M}}$}\fi}

\newcommand{\ha}{H$\alpha$}

\newcommand{\nii}{[N\,{\sc ii}]}

\def \st{\ifmmode{^{\mathrm{st}}}\else{${^{\mathrm{st}}}$}\fi}
\def \nd{\ifmmode{^{\mathrm{nd}}}\else{${^{\mathrm{nd}}}$}\fi}
\def \rd{\ifmmode{^{\mathrm{rd}}}\else{${^{\mathrm{rd}}}$}\fi}
\def \th{\ifmmode{^{\mathrm{th}}}\else{${^{\mathrm{th}}}$}\fi}

\title[NGC 6302]{Hubble-type
outflows of  the high excitation, 
poly--polar planetary nebula NGC 6302 -- from expansion proper motions.}
\author[J. Meaburn et al]
{J. Meaburn$^{1}$, 
M. Lloyd $^{1}$, N. M. H.
Vaytet,  $^{1}$
\& J. A. L\'{o}pez $^{2}$\\
 $^{1}$Jodrell Bank Observatory, Dept of Physics \& Astronomy, University of
Manchester, Macclesfield, Cheshire SK11 9DL UK.\\
$^{2}$
Instituto de Astronom\'{\i}a, UNAM, Apdo. Postal 877,
Ensenada, B.C. 22800, M\'{e}xico.\\
}

\begin{document}

\date{Accepted yyyy mmmmmmmmmm dd. Received yyyy mmmmmmmmmm dd; in original 
form yyyy mmmmmmmmmm dd}

\pagerange{\pageref{firstpage}--\pageref{lastpage}} \pubyear{2004}

\maketitle

\label{firstpage}

\begin{abstract}

The outflowing proper motions of fifteen knots in the dominant
northwestern lobe of the high--excitation poly--polar planetary
nebula NGC~6302 have been determined by comparing their positions 
relative to those of faint stars
in an image
taken at the San Pedro Martir Observatory 
in 2007 to those in a South African Astronomical Observatory archival 
plate obtained by Evans in 1956. The Hubble-type expansion of this 
lobe is now directly confirmed in a model--independent way from these
measurements. Furthermore, an unambigous distance to NGC~6302 of 
1.17 $\pm$ 0.14 kpc is now determined. Also all the velocity vectors
of the fifteen knots (and two others) point back to the central source.
An eruptive event from within the central torus 
$\approx$ 2200 yr previously must have created the
high speed lobes of NGC~6302.

\end{abstract}


\section{Introduction}
The poly--polar planetary nebula
NGC 6302 has always attracted interest (see Meaburn et al
2005 - hereafter Paper 1 - and papers therein)  because it has an extremely 
hot star in its central stellar system and because multiple 
bipolar lobes which emanate from this source 
reach outflow velocities of $\geq$ 600 \kms
(Paper 1).  Furthermore, a surprisingly
massive central CO emitting
torus ($\approx$
2  \msun\ and  expanding at 8 \kms) viewed nearly edge--on, obscures
this central stellar system (Matsuura et al 2005; Peretto et al 2007).
These authors estimate that this was ejected between 7500 yr and 2900 yr
ago. The surface 
temperature of the hot central star has been estimated most recently from FIR 
ISO observations 
as 220,000 K by Wright et al. (2007).
This is the lowest temperature yet estimated for the hydrogen
deficient central star.

The high--speed outflow of the principal northwestern lobe  
is Hubble-type i.e. outflow velocity is proportional
to distance from the source. This  has been conclusively 
illustrated by the kinematic--morphological modelling of the
spatially resolved optical line profile
observations in Paper 1, 
though it was also revealed in a more limited fashion in 
Meaburn \& Walsh (1980). This behaviour alone favours a ballistic
origin for the ejecta i.e. simply the gas more distant from the
source has been ejected at the highest speed. This Hubble--type
assertion would also predict
that the velocity vectors of all of the outflowing material in a lobe 
point 
directly back to their source.

Estimations of the crucial distance to NGC~6302 (see Paper 1 for a summary) 
ranged from 0.15 to
2.4 kpc prior to its expansion proper motion (PM) measurement
of 1.04 $\pm$ 0.16 kpc reported in Paper 1. The latter 
value was based on the
kinematic--morphological modelling of the northwestern lobe combined
with a  measurement of the PM of a single compact  knot apparent between the 
1956 image of Evans (1959) and a comparable image taken in 2001.
The prediction that the dynamical age of the northwestern lobe is
1900 yr is also a natural consequence of this PM measurent. This would suggest
that the lobe was ejected after the central torus had formed.

The original 1956 plate taken of NGC~6302 
by Evans (1959) has now been obtained
from the SAAO archive and scanned to permit comparison  with a new
image of the northwestern lobe taken in 2007. 
The northwestern lobe is particularly suited to this analysis
of its expansion PMs 
for its axis is close to the plane of the sky and its outflow
velocities become high in its extremities and are now well--known (Paper 1).
PMs of fifteen nebular knots distributed along the length of 
the northwestern lobe have now been measured. The confirmation
of the Hubble--type nature of the outflow has now been established
along this lobe
by an independent method as a consequence,
as has the ballistic nature of the ejection process. More refined
values for the expansion--PM distance to NGC 6302 and the dynamical age
of the northwestern lobe have also been derived.

\section{Observations and Results}

The baseline photographic  
plate for the present proper motion measurements was taken with
 with the 74-inch Radcliffe reflector by Evans (1959)
on August 1 1956. This was through a Wratten I205 dye filter which
when combined with Kodak 103aE photographic emulsion has
a bandwidth of a few hundred angstroms centred on the
\ha\ and \nii\ nebular emission lines. The exposure time of 45 minutes
left the image of the bright nebular core saturated but revealed 
most of the northwestern lobe of NGC~6302 (Fig. 1).

The second image of the northwestern lobe (Fig 2)
was taken on June 18 2007 with the Manchester
Echelle spectrometer (Meaburn et al 2003) in its imaging mode when
combined with the 2.1--metre San Pedro Martir telescope. A 100\AA\ bandwidth
interference filter isolated the \ha\ and \nii\ lines and a SITE CCD
was the detector. Two times binning made each of the 512$\times$512
data taking windows 0.6224\arcsec\ across when projected on to the sky.
The exposure time was 60 seconds. The time separation between
the two images is then 50.88 yr. The coordinates for the images in
Fig. 1 \& 2 were derived from stellar positions taken from
the POSS digital sky survey and, because of their limited
accuracy  (a few arcsec) are useful only to identify nebular features.

The 1956 plate was scanned by a transmission, slide scanner to convert
the photographic density distribution into a digital array. The digital
values in the subsequent array have a complex relationship to 
the relative surface brightnesses of nebular features; whereas 
the values within the 2007 CCD image are  nearly linearly related. However,
the positions of the brighteness maxima of compact nebular knots, 
a few arcsec across, are closely comparable in both arrays. Fifteen
such knots (1--15 in Table 1) in unsaturated areas of the 1956 image
of the northwestern lobe of NGC 6302
were identified. Two (16 \& 17) are outside
this lobe.

A subset of the 1956 data array (Fig. 1), matching the coverage of the
northwesterm lobe by the 2007 imagery (Fig. 2), was obtained using the ISUBSET
FIGARO  software routine and scaled to the same pixel size
as the  2007 array using the SQORST KAPPA routine. The separations of
stellar images in both arrays were used to derive the scaling factor.
Sections of both arrays were then contoured where each section 
contains some of the  
knots as well as sufficient images of the faint stars.
Examples of the resultant contour maps from this process are shown
in Figs. 3a-- b for the area marked by a rectangle in Fig. 1.

The shifts of the nebular knots 1--7 listed in Table 1 and identified
in Fig. 4 over the 50.88 yr base line are immediately apparent when
the positions of their brightness maxima are compared to those of the
stellar images in Figs. 3a \& b. The faint stars with
PMs $\leq$ 0.3\arcsec\ over the same baseline are easily identified
by measuring their image positions relative to the other faint
star positions in the same array. For example, of the five faint stellar
images common to both contour maps in Fig. 3a--b (marked S1,S3-6) 
only that marked S4, to
the bottom left of both arrays, shifted by this amount over the 50.88 yr
baseline (the bright star also marked S2 to the top of the array 
was considered unreliable for this purpose). The differences in the 
positions of the knots 1--7 were then measured by comparing their
shifted positions  to the unshifted mean
positions of the four remaining faint stars in this example i.e.
S1, S3, S5 \& S6. Three 
further contour maps covering knots 8--17 were used in an identical manner.
The results of this process are listed in Table 1 and illustrated
graphically in Fig. 5. Repeated measurements of the
contour maps illustrated that the  shift in the position of each knot
had been derived to $\pm$ 0.3\arcsec\ accuracy which gives rise to
the error bar for all the PM measurements of 6 mas yr$^{-1}$ in Fig. 5
where these PMs are plotted versus separation from the central thermal 
radio source (Terzian, Balick \& Bignell 1974) and infrared point
source (Danziger et al 1973). This position (see Table 1) is
also the centroid of the CO map of the core of NGC~6302 (Peretto et al 2007).
The measurements of the 
directions of the motions of most of the knots by the same process
are accurate to $\pm$ 2\degree\ except for the two elongated knots
(10 and 17 in Fig.4) where the uncertainty is a few times this amount. 

\section{Discussion}

\subsection{Hubble--type expansion}

In Paper 1 a Hubble--type expansion of the northwestern lobe
of NGC~6302 was indicated after the observed radial velocities 
were compared with the predictions of a  morphological-kinematical
model. The starting point for this modelling came simply from
the pv array (fig. 4 of Meaburn \& Walsh 1980) across the lobe at 
1.71\arcmin\ from the central source. This revealed that the 
axis of the lobe is 
inclined at only $\phi$ = 12.8\degree\ $\pm$ 2\degree\ to the plane of the sky 
with its nearside
edge flowing away from the source nearly in the plane the sky
(see fig 13 Paper 1). An outflow
velocity of 263 $\pm$ 20 \kms\ was then derived at this position.
From this starting point the 
more extensive motions over the northwestern lobe in
Paper 1 matched closely the 
model predictions where 
a Hubble--type expansion, with the motions of all parts of the lobe
directed away from the central source, are both assumed.

The measurements of the PMs  and their flow directions 
of 15 nebular knots in the northwestern lobe 
therefore permit the validity of 
both of these predictions to be examined directly: the 
fortuitous small value of $\phi$ 
and the
high elongation of the lobe mean that no modelling is necessary
in the interpretation of the results. The linear increase in PM with
distance along the northwestern lobe, as expected for
Hubble--type motions, is clearly shown in Fig. 5. Moreover the velocity
vectors in Fig. 4 are all directed away from the central source
at RA(2000) 17:13:44.48, DEC(2000) -37:06:14.7 (Table 1) within
the measurement uncertaintities of $\pm$ 2\degree\ for each vector. 

The conclusions of Paper 1 are therefore directly confirmed: 
the knots were ejected in the same short--lived eruptive event
and have travelled outwards ballistically for around 2200 $\pm$ 100 yr (the
weighted mean of the very similar PM dynamic ages given in Table 1); with the
fastest travelling proportionally furthest.
The larger mean dynamical age of the knots 16 and 17 (see Table 1
and Fig. 4) could marginally
indicate that these were emitted at a somewhat earlier time
than the dominant northwestern lobe.
 
\subsection{Distance}

The most accurate distance D (kpc) to NGC 6302 up to that date 
is derived in Paper 1 as 1.04 $\pm$ 0.16 kpc using the PM of Knot 2
(Fig. 4) whose 1956 position was measured with a ruler 
off an enlargement of 
the published figure
in Evans (1959).
The present work has improved the measurement of the PM 
of Knot 2 and now determined those of 14 other knots in the northwestern lobe
(Table 1).There was always the possibility that deriving D from
one knot alone as in Paper 1 would give an anomolous answer.
 

The relationship of D, PM and outflow velocity V along a line at an angle
$\phi$ to the plane of the sky is:
{\center D(kpc) $\times$ PM(mas y$^{-1}$) = 0.2168 $\times$ V (\kms) $\times$
cos($\phi$) \hspace{3cm}  (1)}

and for V = 263 \kms, at 1.71\arcmin\ from the central source and
for $\phi$ = 12.8\degree\ (for the axis of the northwestern
lobe). Assuming a Hubble--type outflow then for a knot at 
{\bf \it x}(arcmin) from the central source (column 5 in Table 1),
{\center 
D(kpc) = 32.51 $\times$ {\bf \it x}(arcmin) $\times$ 
(PM(mas y$^{-1}$))$^{-1}$. \hspace{4cm}         (2)}

The values derived for D of each knot (column 7 in Table 1)
are for the angular separations given in column 5 combined with the PMs in
column 3 derived with equ. 2. The random 
errors in column 7 are a consequence only of the $\pm$ 6 mas y$^{-1}$
uncertainty
on each PM value in column 3.
The weighted mean of D from the values in column 7 is therefore 
1.17  kpc. The uncertainty in this 
weighted mean is $\pm$ 0.03 kpc if only the 
standard deviation of the error values in column 7 of Table 1 are taken
into account. The actual weighted mean standard deviation of the
values for D in column 7 of Table 1 is   $\pm$ 0.09 kpc.
This discrepency indicates that either the knots themselves have
a significant dispersion of outflowing velocities around the Hubble--type
prediction or that a further source of random uncertainty is present. 
The most likely origin for the latter is that 
 $\phi$ for an individual knot can range
from $\pm$ 12.8\degree\ around the lobe axis at  $\phi$ = 12.8\degree\
(see Paper 1)
then cos($\phi$) in equ. 1 can have values from 1 to 0.9 depending
whether or not a knot is towards the nearside or farside of the
northwestern lobe, which cannot be determined in the present observations.
The actual 
standard deviation around the weighted mean value 
of $\approx \pm$ 0.1 kpc of the D values in Table 1
which were derived assuming all the knots had the same $\phi$ then most likely
 reflects
this angular spread of the outflow.  
In addition, a systematic 
uncertainty is present  in the mean value of D 
due to the use of V = 263 $\pm$ 20 \kms\
at  {\bf \it x} = 1.71\arcmin\ from Paper 1 in equ. 2 therefore
D = 1.17 $\pm$ 0.14 kpc
is a realistic best value, with a conservative estimation of the uncertainty, 
for the distance to NGC~6302.

Since there is excellent evidence of both velocity increasing linearly 
with distance  
(Hubble law expansion) and PM vectors pointing back to the central source,  
and therefore an eruptive event as origin of lobes, then fragmentation  
of the eruption that is channeled perpendicular to the thick torus   
explains all of above plus the formation of the poly-polar or multi- 
polar structures in a natural way, i.e. all `lobes' are formed at the  
same time through the splitting or fragmentation of the mass  
participating in the eruptive event. Evidence of this fragmented mass  
is clearly appreciated in the HST images of the core of
NGC~6302 presented by Matsuura et al  
(2005). 
Furthermore, the outflowing velocities of $\geq$ 600 \kms\ (Paper 1) for the
knotty extremities of the northwestern lobe of NGC~6302 are similar
to those of the knots in the Hubble-type outflow of the Hourglass
Nebula MyCn~18 (Bryce et al 1997; O'Connor et al, 2000). Similar
eruptive
processes must have been in play.
Steffen \& L\'opez (2004) demonstrated theoretically that some of these
properties could be generated by a fast wind blowing through a clumpy
medium. This however, is an unlikely mechanism for the creation
of the NGC~6302 lobes for there is no direct evidence of a fast
wind (Paper 1), the fastest knots are the densest 
(1000 cm$^{-3}$ - Meaburn \& Walsh, 1980) contrary to their
predictions and the $\geq$ 600 \kms\ outflow speeds are too high.
More probably the present outflow is the consequence of a nova--type
explosion or an eruputive event channeled down the rotation
axis of a close binary system.

\section{Conclusions}

A Hubble--type outflow has been shown to be occurring 
directly from PM measurements
of 15 knots in the northwestern lobe of NGC~6302. This confirms the 
model--dependent prediction of the same behaviour 
from radial velocity measurements in Paper 1 (and see Corradi 2004
for similar behaviour in other PNe).

The velocity vectors of these 15 knots in NGC ~6302 
point back to the central source.

The outflowing velocities of $\geq$ 600 \kms\ at the furthest extremities
of the northwestern lobe are confirmed.

The northwestern lobe of NGC~6302 must then have been created in an
eruptive event $\approx$ 2200 yr ago but probably after the creation of the
central torus 2900 - 7500 yr ago as estimated by 
Matsuura et al (2005) \& Peretto et al (2007).

The distance to NGC~6302 is now determined unambiguously  
as D = 1.17 $\pm$ 0.14 kpc.

 \section*{Acknowledgements}
JM is grateful to UNAM for supporting his 2007 visit to the Instituto
de Astronom\'{i}a, Ensenada where this paper was initiated.
ML (n\'{e}e Bryce) and  and  NMHV thank the staff at the SPM observatory
in 2007 for their help during the latest imagery. JM thanks T. Lloyd--Evans
for bringing the 1956 SAAO archive plate to the UK.
JAL 
is in grateful receipt of DGAPA - UNAM grants IN 112103 \& 108406-2
and NMHV a Manchester University School of Physics and Astronomy bursary.

\begin{table*}
\centering
\caption{Column 1 gives knot identication from Fig.4. 
Knots 1--15 are in the northwestern lobe.
Column 2 gives
the RA and DEC (2000 coords) from the 2007 image. These are for
identification purposes only and are not accurate enough ($\pm$ 1\arcsec)
for future proper motion measurements.Column 3 gives the PM of each knot
derived from the displacement between its 1956 and 2007 positions.
All values are derived conservatively to $\pm$ 6 mas yr$^{-1}$ accuracy.Column
4 contains the position angles of the direction of motion between these dates
to  $\pm$ 2\degree\ accuracy for all but 10 and 17 (see text).
Column 5 gives the separation of each knot from the central 
compact radio source
(RA(2000) 17:13:44.48, DEC(2000) -37:06:14.7 from Terzian, Balick 
\& Bignell 1974).
Column 6 gives the proper--motion dynamical age of each knot. Column 7
gives the distance to NGC 6302 derived from the values in columns 3 \& 5
with the uncertainties derived from the $\pm$ 6 mas yr$^{-1}$ in the PM
values in column 3.}

\begin{tabular}{|l|c|c|c|c|c|c|}

\hline
\
{1. \bf Knot}&{2. \bf Position}&{3. \bf PM}&{4. \bf PA}&{5. \bf 
Angular separation }&{6. \bf Dynamic age}&7. \bf Distance\\
{}&{\bf RA,DEC(2000) }&{\bf mas yr{$^{-1}$}}&{\bf deg}&{\bf arcmin}&{\bf yr}&
kpc\\
1)&17:13:30.66, -37:05:22.2&78.6&287&2.95&2250&1.22 $\pm$ 0.09\\
2)&17:13:34.15, -37:05:15.2&68.7&304&2.28&1990&1.08 $\pm$ 0.09\\
3)&17:13:31.29, -37:05:17.9&64.9&289&2.06&1910&1.03 $\pm$ 0.10\\
4)&17:13:31.07, -37:05:30.3&74.7&283&2.84&2280&1.24 $\pm$ 0.10\\
5)&17:13:32.38, -37:05:18.7&70.8&295&2.58&2190&1.23 $\pm$ 0.10\\
6)&17:13:32.84, -37:05:25.0&64.9&289&2.46&2280&1.23 $\pm$ 0.11\\
7)&17:13:32.89, -37:05:38.7&66.8&289&2.39&2150&1.16 $\pm$ 0.11\\
8)&17:13:33.72, -37:05:44.4&55.0&289&2.21&2410&1.31 $\pm$ 0.14\\
9)&17:13:34.71, -37:05:38.9&59.0&303&2.04&2080&1.12 $\pm$ 0.11\\
10)&17:13:35.38, -37:05:47.1&55.0&280&1.94&2110&1.15 $\pm$ 0.13\\
11)&17:13:35.53, -37:05:57.0&49.1&283&1.81&2210&1.20 $\pm$ 0.15\\ 
12)&17:13:36.41, -37:06:00.9&39.3&264&1.63&2480&1.35 $\pm$ 0.21\\
13)&17:13:38.91, -37:06:03.7&25.6&273&1.12&2630&1.42 $\pm$ 0.33\\
14)&17:13:40.01, -37:05:47.6&29.5&292&1.06&2160&1.26 $\pm$ 0.26\\
15)&17:13:39.86, -37:05:35.8&35.4&306&1.07&1820&0.98 $\pm$ 0.17\\
\hline
16)&17:13:33.23, -37:06:28.5&39.3&271&2.26&3450\\
17)&17:13:33.49, -37:06:24.8&55.0&256&2.20&2400\\
\hline
\end{tabular}
\end{table*}

\clearpage

\begin{figure*}
\epsfclipon
\centering
\mbox{\epsfysize=4in\epsfbox[0 0  280 200]{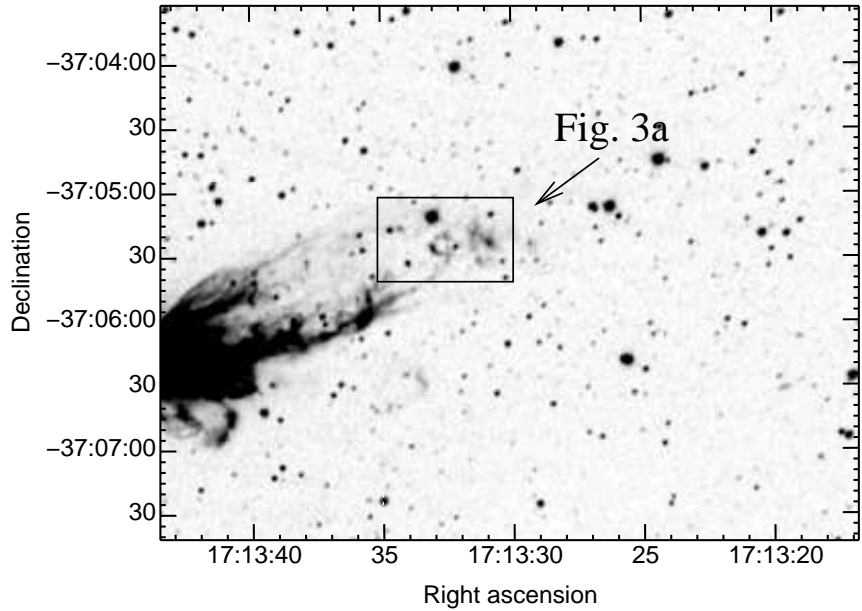}}
\caption{A section from the red 1956 photographic plate taken by Evans (1959).
This is a subset covering only the northwestern lobe of NGC~6302.
The nebular lines \ha\ and \nii\ were isolated by the bandwidth.
The rectangle identifies the area contained within the contour map
in Fig. 3a}
\label{reffig1}
\end{figure*}

\begin{figure*}
\epsfclipon
\centering
\mbox{\epsfxsize=5in\epsfbox[96 257 443 511]{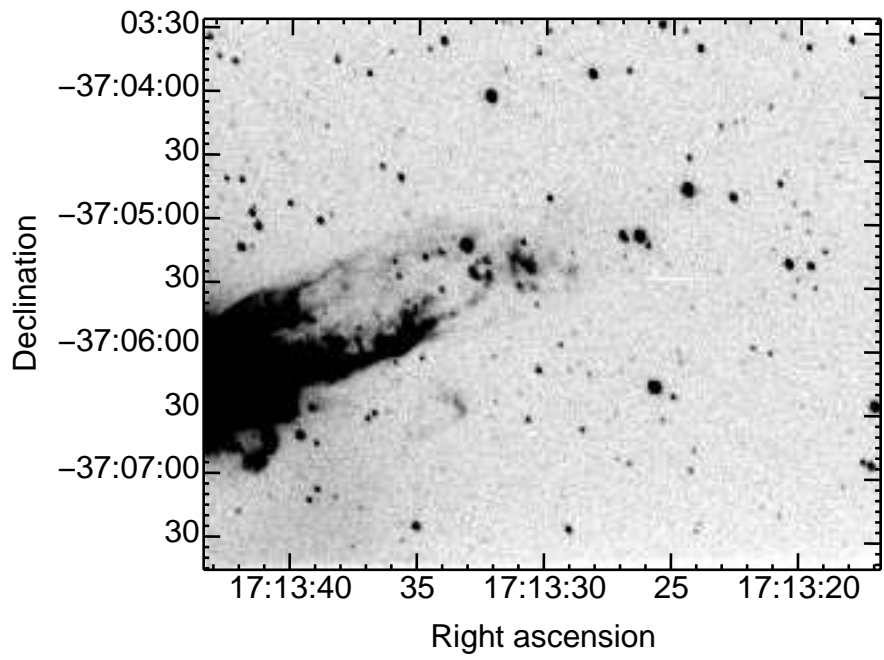}}
\caption{The same area of NGC~6302 as shown in Fig. 1 but
obtained in 2007. The same nebular lines were transmitted.}
\label{reffig2}
\end{figure*}

\begin{figure*}
\epsfclipon
\centering
\mbox{\epsfysize=6in\epsfbox[0 35 505 765]{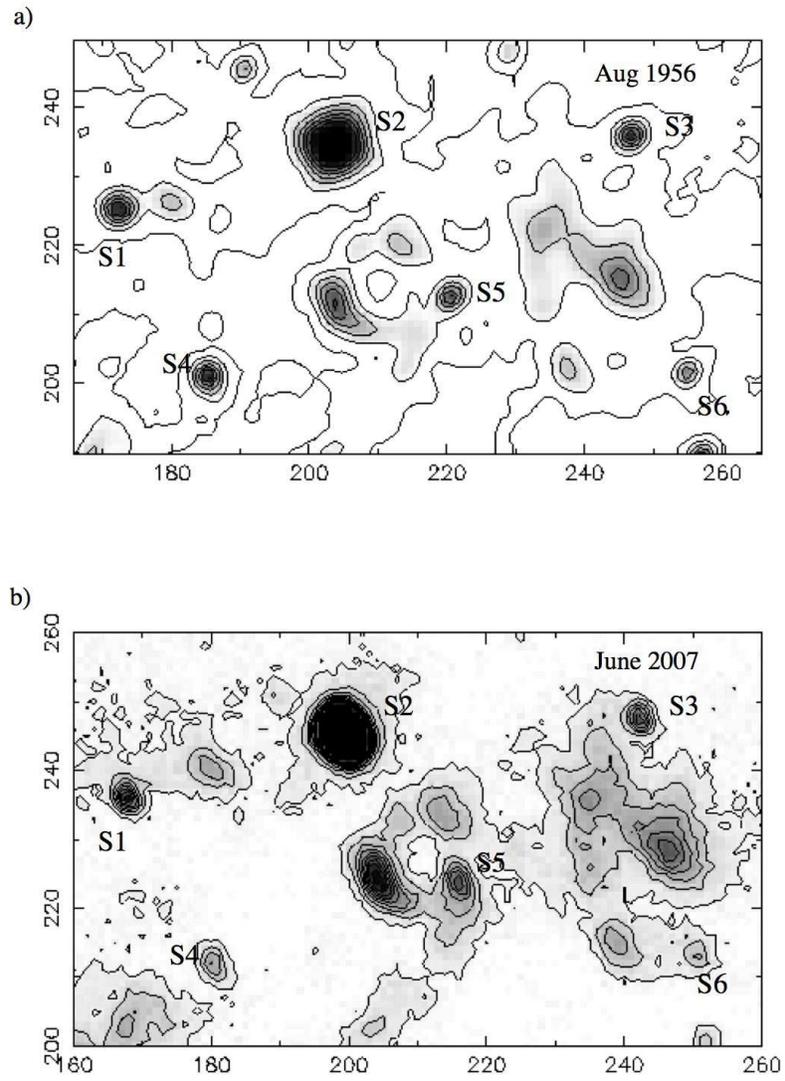}}
\caption{a) A contour map and negative greyscale presentation of 
the subset of the 1956 plate whose area is identified in Fig. 1.
Star images are identified by 'S1--6` and the contours
of knots 1--6 can be seen and identified from Fig. 4. b) As for Fig. 3a 
but for the CCD image taken in 2007.}
\label{reffig3}
\end{figure*}

\begin{figure*}
\epsfclipon
\centering
\mbox{\epsfysize=4in\epsfbox[6 166 464 554]{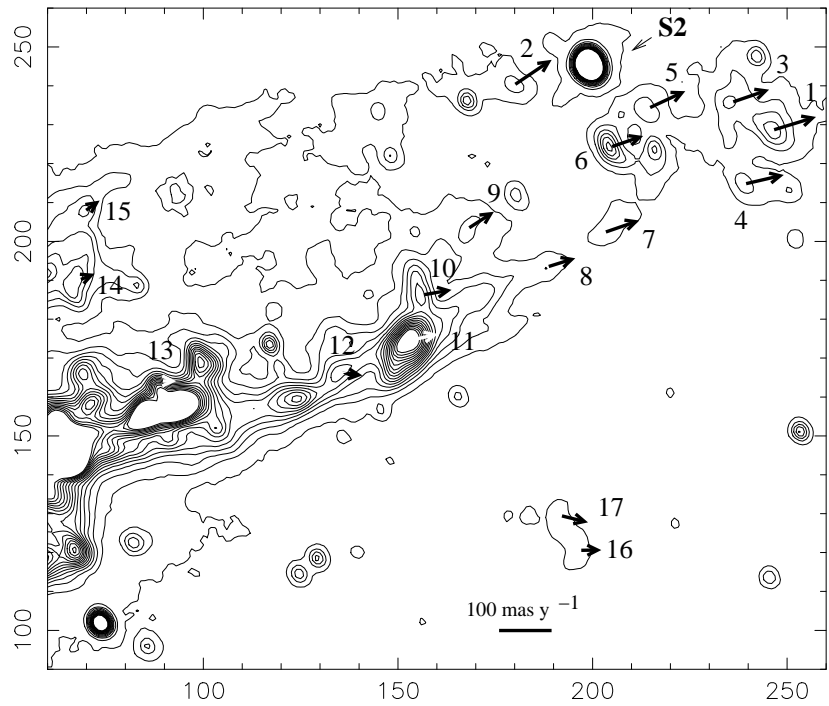}}
\caption{A contour map is presented of that part of the northwestern
lobe in the 2007 CCD array 
that contains the knots 1--15 and the southern feature
knots 16--17. The expansion PM vectors are taken from their
values in Table 1. The extent of the displacement is from the
tip of each arrow to the end of its tail. Note that the arrows 
for knots 11 and 13 are white to show against packed contour lines.
The star S2 from Figs. 3a \& b is identified to aid the comparisons
between these contour maps.}
\label{reffig4}
\end{figure*}

\begin{figure*}
\epsfclipon
\centering
\mbox{\epsfysize=4in\epsfbox[0 0 523 598]{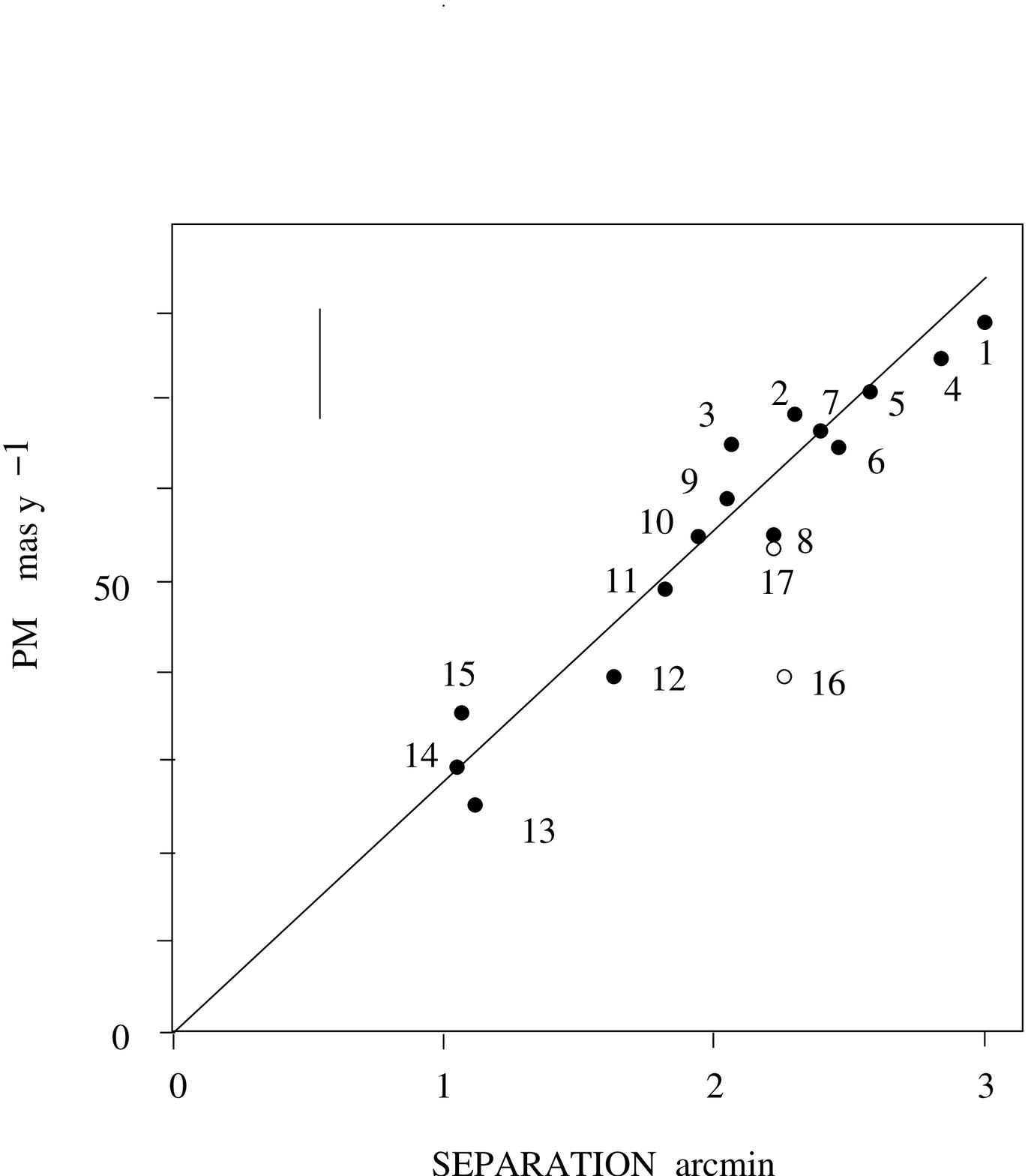}}
\caption{The PMs of knots 1--17 are shown against their separation from
the obscured, central, source. The errors in each displacement
$\pm$ 0.3\arcsec\ for each knot over the 50.88 yr baseline is nearly the same
and, because of the Hubble-type expansion translates to an error
of $\pm$ 6 mas yr$^{-1}$ for all of the proper motions. Knots 16 and 17
are shown as white circles for the do not belong to the northwestern
lobe. A weighted mean, least 
squares  best fit straight line is shown for the northwestern
lobe knots only.}
\label{reffig5}
\end{figure*}

\label{lastpage}
\end{document}